\documentclass[a4paper]{jpconf}
\usepackage{graphicx}
\usepackage{citesort}
\begin{document}
\title{Synergy of multifrequency studies from observations of NGC6334I}

\author{Andreas Seifahrt$^1$, Sven Thorwirth$^2$, Henrik Beuther$^3$, Silvia Leurini$^4$, Crystal L Brogan$^5$, Todd R Hunter$^5$, Karl M Menten$^2$ and Bringfried Stecklum$^6$}

\address{$^1$ Georg-August-Universit\"at, Institut f\"ur Astrophysik, Friedrich-Hund-Platz 1, 37077 G\"ottingen, Germany}
\address{$^2$ Max-Planck-Institut f\"ur Radioastronomie, Auf dem H\"ugel 69, 53121 Bonn, Germany}
\address{$^3$ Max-Planck-Institut f\"ur Astronomie, K\"onigstuhl 17, 69117 Heidelberg, Germany}
\address{$^4$ European Southern Observatory, Karl-Schwarzschild-Strasse 2, 85748 Garching, Germany}
\address{$^5$ NRAO, 520 Edgemont Rd, Charlottesville, VA 22903, USA}
\address{$^6$ Th\"uringer Landessternwarte Tautenburg, Sternwarte 5, 07778 Tautenburg, Germany}

\ead{seifahrt@astro.physik.uni-goettingen.de}

\begin{abstract}
We combine multifrequency observations from the millimeter to near
infrared wavelengths that demonstrate the spatial distributions of
H$_2$, CO, and NH$_3$ emission, which are all manifestations of various
shocks driven by outflows of deeply embedded sources(s) in NGC6334I.
In addition to the well-known northeast-southwest outflow we detect at
least one more outflow in the region by combining observations from
APEX, ATCA, SMA, \textit{Spitzer} and VLT/ISAAC. Potential driving sources will be
discussed. NGC6334I exhibits several signs of active star formation and
will be a major target for future observatories such as \textit{Herschel} and ALMA.
\end{abstract}

\section{Multifrequency studies in NGC6334I}

NGC6334 is a giant molecular cloud located at a distance of 1.7 kpc \cite{neckel78} in the southern galactic plane. Along a gas and dust filament of 11 pc, NGC6334 exhibits several luminous sites of massive star formation – as seen in the far-infrared \cite{mcbreen79} and radio continuum \cite{rodriguez82}. Emission from the sub-site NGC6334I dominates the millimeter to the far infrared region \cite{sandell00}. A near-infrared image of the NGC6334 is shown in Fig. 1, where we 
combine the \textit{Spitzer} IRAC channels to a colour composite. 

Single dish molecular line observations show NGC6334I to be chemically rich, comparable in line density (and hence chemical complexity) to prototypical hot cores such as Orion-KL and SgrB2(N) (e.g., \cite{sven03,schilke06}). ATCA investigations of NH$_3$ emission up to the (6,6) inversion transition reveal the presence of warm gas \cite{beuther07}. SubMillimeter Array (SMA) continuum observations at 1.3~mm \cite{hunter06} resolve NGC6334I into a sample of sub-cores of several tens of solar masses each, nicely demonstrating the formation of star clusters. 

NGC6334I has also been observed in the infrared \cite{persi05}.
In recent years, with the advent of the new generation of large optical/infrared telescopes of the 8-10 m class, several pioneering studies have been performed. Mid-infrared CTIO and Keck II imaging of NGC6334I resolving the central UCHII region into two distinct sources was reported \cite{debuizer00}. Magellan-Clay PANIC near-IR ($JHKs$) observations identified a high-mass young star, IRS1-E, as the powering source of the UCHII region \cite{persi05}.

Given its rich observational history, NGC6334I is on its way of becoming one of the very few templates of high-mass star formation in the entire sky. It will also be a major target for future observatories such as \textit{Herschel} and ALMA. 

We demonstrate here the synergy obtained from combining published studies from multifrequency observations with a so far unpublished high resolution VLT/ISAAC image of shocked H$_2$ gas surrounding the central source of NGC6334I (see Fig. 2). 
We identify five knots in the immediate vicinity of NGC6334I and denote them by  $\alpha$, $\beta$, $\gamma$, $\delta$, and $\epsilon$ (see Fig. 3). While three out of these five H$_2$ knots were previously known \cite{persi96}, two new knots ($\gamma$ and $\delta$) could be identified. Their counterparts may lie well hidden behind the UCHII region (see Fig. 4). 
Two of the previously known knots ($\alpha$ and $\beta$) coinside nicely with CO (4-3) emission lobes detected with APEX \cite{leurini06} (see Fig. 5). At least one of the new H$_2$ knots coincides also with NH$_3$ maser emission (see Fig. 6), as is the case for the previously known knots, demonstrating the power of combining multifrequency observations to reveal the morphology of outflows from deeply embedded sources. 

\section*{References}
\bibliography{seifahrt}



\newpage

\begin{figure}[t]
\begin{center}
\includegraphics[bb=50 40 660 570,clip,width=10.5cm]{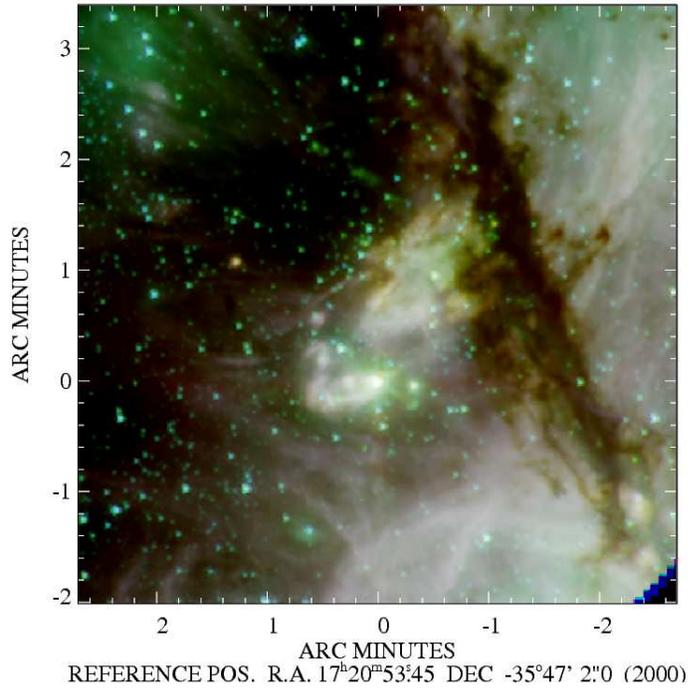}\hspace{2pc}%
\end{center}
\begin{minipage}[b]{16cm}\caption{\label{fig1}Colour composite from \textit{Spitzer} IRAC observations (P.I. Fazio, 'Deep IRAC Imaging of High Mass Protostars') in 3.6$\mu$m (blue), 4.5$\mu$m (green) and 5.8$\mu$m (red). Data source: SPITZER archive facility.}
\end{minipage}
\end{figure}

\begin{figure}[b]
\begin{center}
\includegraphics[bb=50 40 640 570,clip,width=11cm]{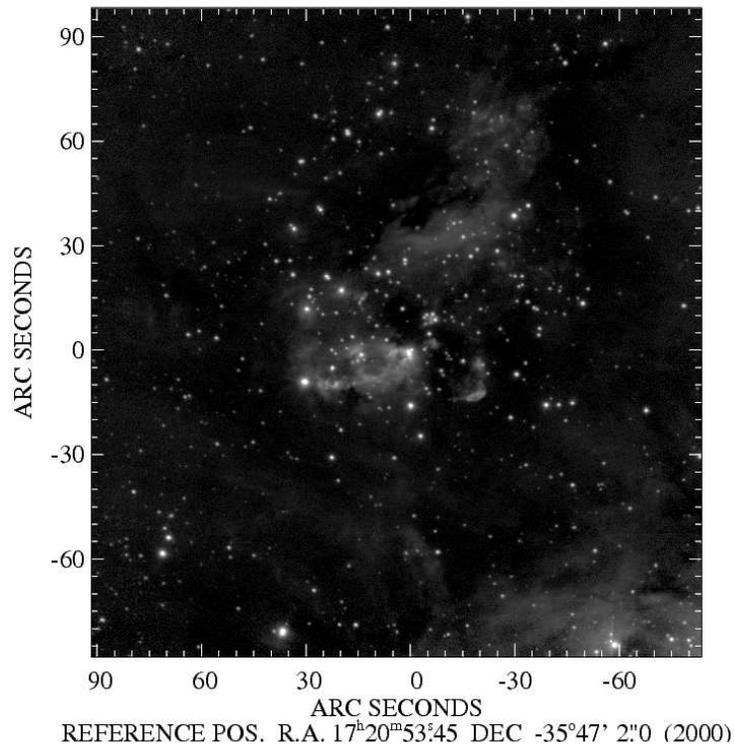}\hspace{2pc}%
\end{center}
\begin{minipage}[b]{16cm}\caption{\label{fig2}VLT/ISAAC observations in H$_2$ (2.12$\mu$m) and Br$\gamma$ (2.17$\mu$m) combined to one image. The spatial resolution is $\sim$0.5 arcsec. }
\end{minipage}
\end{figure}

\begin{figure}[t]
\begin{center}
\includegraphics[bb=50 40 660 570,clip,width=9.5cm]{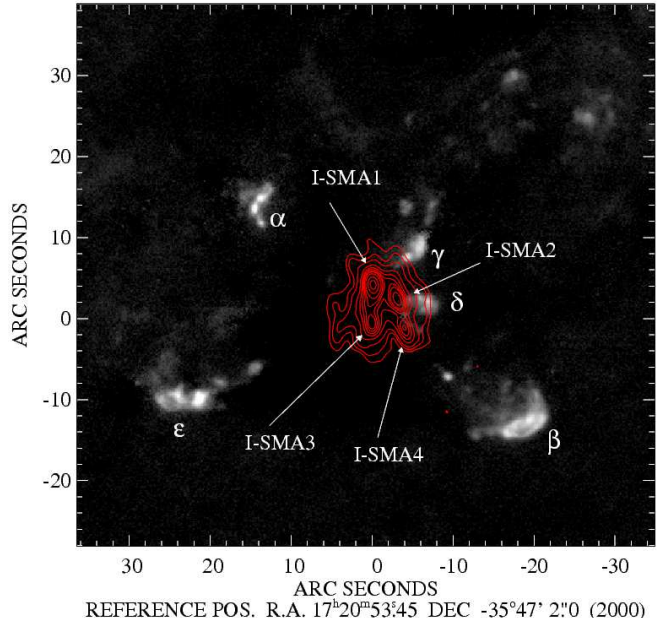}\hspace{2pc}%
\end{center}
\begin{minipage}[b]{16cm}\caption{\label{fig3} The composite, obtained by subtracting the VLT/ISAAC 2.17$\mu$m image from the 2.12$\mu$m image (see Fig. 2), shows five H$_2$ knots in the immediate vicinity of NGC6334I. We denote these knots by $\alpha$, $\beta$, $\gamma$, $\delta$, and $\epsilon$. Overplotted as red contours are SMA observations from \cite{hunter06} showing four 1.3 mm continuum sources, denoted by I-SMA 1-4.}
\end{minipage}
\end{figure}

\begin{figure}[b]
\begin{center}
\includegraphics[bb=50 40 620 620,clip,width=9.5cm,height=9cm]{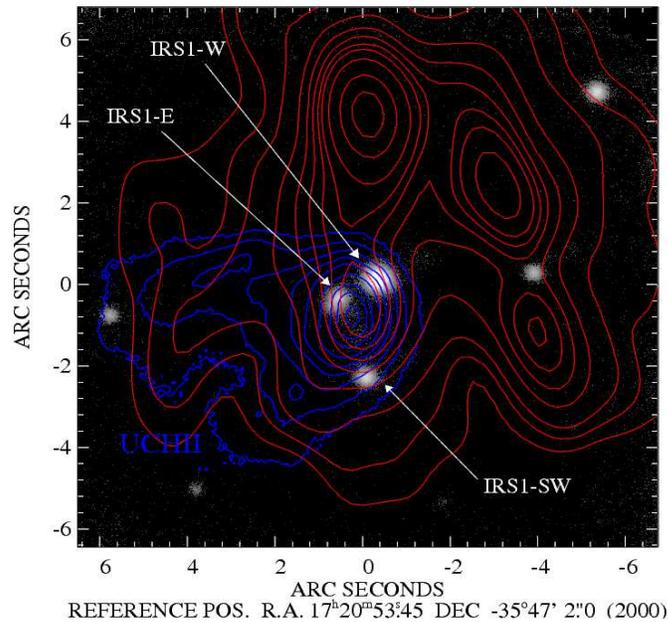}\hspace{2pc}%
\end{center}
\begin{minipage}[b]{16cm}\caption{\label{fig4}VLT/NACO H-band image, overplotted with the 1.3 mm contours of \cite{hunter06}  (red contours) and 3$\mu$m continuum emission from a NACO L-band image (blue contours). Three point sources in the NACO image coincide with the 1.3mm source I-SMA3. We adopt the nomenclature of \cite{persi05}. With a morphology similar to the centimeter continuum, the L-band emission traces the UCHII region, which might be driven by the star IRS1-E.}
\end{minipage}
\end{figure}

\begin{figure}[t]
\begin{center}
\includegraphics[bb=50 40 660 570,clip,width=10cm]{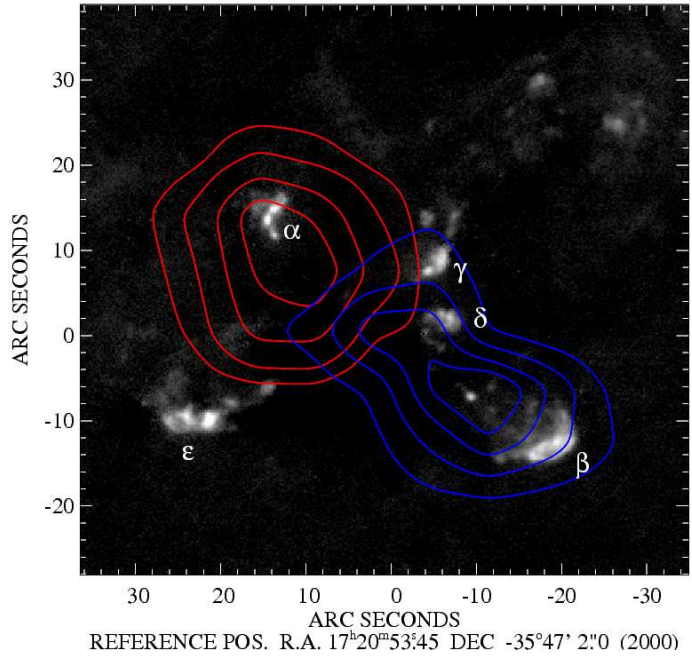}\hspace{2pc}%
\end{center}
\begin{minipage}[b]{16cm}\caption{\label{fig5}VLT/ISAAC H$_2$ (2.12$\mu$m) emission. Overplotted are APEX observations from \cite{leurini06}, showing CO (4-3) emission in the +7 to +55 km/s (red) and $-78$ to $-15$ km/s (blue) ranges. Both CO lobes coincide well with the H$_2$ lobes $\alpha$ and $\beta$.}
\end{minipage}
\end{figure}

\begin{figure}[b]
\begin{center}
\includegraphics[bb=50 40 660 570,clip,width=10cm]{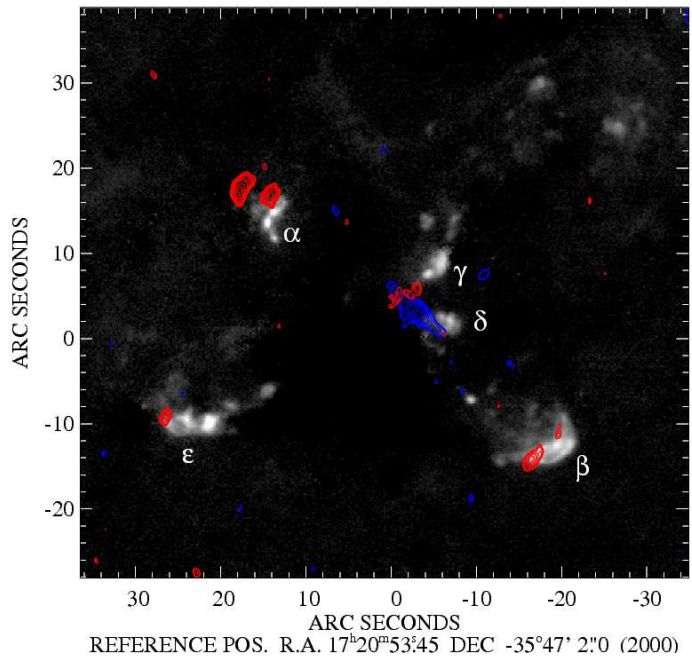}\hspace{2pc}%
\end{center}
\begin{minipage}[b]{16cm}\caption{\label{fig6}VLT/ISAAC H$_2$ (2.12$\mu$m) emission. Overplotted are ATCA observations from \cite{beuther08} showing NH$_3$ (3,3) and (6,6) maser emission (red and blue contours, respectively). Note the correlation between the (3,3) emission and the H$_2$ lobes $\alpha$, $\beta$, and $\epsilon$. The central NH3 (3,3) emission feature seems to connect the H$_2$ lobe $\gamma$ with the 1.3 mm source I-SMA I (see Fig. 3).}
\end{minipage}
\end{figure}

\end{document}